\documentclass[fleqn,12pt,draft]{article} %%% -> preprint works !
\usepackage{amstex}
\usepackage{righttag}
\usepackage{ctagsplt}
\usepackage{intlim}
\def\oo{\infty}

\def\NuPh{{\it Nucl. Phys. }}
\def\PL{{\it Phys. Lett. }}
\def\PR{{\it Phys. Rev. }}

\def\IJMP{{\it Int. J. Mod. Phys. }}

\def\NK{{N_{k}}} % numero loops
\def\ND{{N_{d}}} % numero denominatori
\def\IOMOG{I^{HO}}
\def\INOMOG{I^{NH}}
\def\dk#1{[d^Dk_{#1}]}
\def\e{\epsilon}
\def\fe{\phantom{\epsilon}}

\def\K{K}
\def\R{R}
\def\C{\eta}
\def\sD{\bar s}
\def\RHO{\boldsymbol{\rho}}
\def\PI{\boldsymbol{\pi}}
\def\Gammae{\Gamma_\e}
\def\SYS{{\tt SYS}}
\def\TITLE{Calculation of master integrals by difference equations}
\def\IDENTIFY{\par (S. Laporta, \TITLE)}
\def\CAPTIONFIG{Diagrams corresponding to the master integrals evaluated.}
\def\CAPTIONTABLE{Values of $J(x)$, $\C_1 I_1(x)$, $I_3(x)$ and $I(x)$, and number
of terms of the series.}
\def\TableOne{
\begin{table}
\footnotesize
\begin{center}
\begin{tabular}{rrrrrr}
\hline
$x$& $J(x)\qquad$ &$\C_1 I_1(x)\qquad$ & $I_3(x)\qquad$& $I(x)\qquad$  & terms \\ \hline
30 & $0.00123 +0.00408\e$  &$0.00578 +0.01918\e$ &$-0.00061 
-0.00141\e$ &$0.00517 +0.01776\e$ &$96$ \\
10 & $0.01388 +0.02799\e$  &$0.03511 +0.07066\e$ &$ -0.00694
-0.00673\e$ &$ 0.02816 +0.06392\e$& $1459$ \\
8  & $0.02380 +0.04062 \e$ &$  0.05229       +0.08892  \e$&   $ -0.01190
-0.00766 \e$&$0.04038 +0.08125\e$  &$5865$\\
7  & $0.03333 +0.05020 \e$ &$  0.06700       +0.10036 \e $&   $ -0.01666
-0.00715 \e$&$ 0.05034 +0.09320\e$ &$15950$\\
6  & $0.05     +0.06280 \e$ &$ 0.09026       +0.11219 \e $&   $ -0.025
-0.00392 \e$& $0.06526 +0.10826\e$ &$73000$\\
5  & $0.08333 +0.07689 \e$ &$  0.13123       +0.11793 \e $&   $ -0.04166
+0.00901 \e$& $0.08957 +0.12694\e$&$661000$ \\
4  & $0.16666 +0.07046 \e$ &$  0.21768       +0.07976 \e $&   $ -0.08333
+0.06720 \e$& $0.13435 +0.14697\e$&$>10^6$ \\
3  & $0.5      -0.28860 \e$ &$ 0.48639       -0.39182 \e $&   $ -0.25
+0.53495 \e$& $0.23639 +0.14313\e$&$>10^6$\\
2  & $\e^{-1}  -0.57721\fe$ &$  0.25 \e^{-1} +0.46029 \fe$&   $
-0.25\e^{-1}+0.14430 \fe$& $0.60459 -0.23190 \e$& \\
1  & $-\e^{-1} -0.42278\fe$ &$  0.25 \e^{-1} -1.45810 \fe$&   $
0.75\e^{-1}+1.06708 \fe$&$ \e^{-1} -0.39101\fe$& \\
0  & $0\phantom{.12345\fe}$ &$ -0.5 \e^{-1}  -0.21139 \fe$&   $ -0.5\e^{-1}
-0.21139 \fe$& $-\e^{-1}-0.42278\fe$ & \\
\hline
\end{tabular}
\end{center}
\caption{\CAPTIONTABLE}
\label{tableij}
\end{table}
}
\def\FigureOne
{
\begin{figure}
\begin{center}
 \begin{picture}(370,180)(0,0)
 \thicklines
\put( 40,100){\makebox(0,0)[lb]{
   \put(000,000){\circle{40}}
   \put(000,000){\line(0,1){20}}
   \put(000,000){\line(+5,-3){17}}
   \put(000,000){\line(-5,-3){17}}
   \put(-11,-40){\text{\LARGE(a)}}
}}
\put(110,100){\makebox(0,0)[lb]{
   \put(000,000){\circle{40}}
   \put(000,-20){\line(0,+1){40}}
   \put(-20,000){\line(-1,0){10}}
   \put(+20,000){\line(+1,0){10}}
   \put(-12,-40){\text{\LARGE(b)}}
}}
\put(180,100){\makebox(0,0)[lb]{
   \put(000,000){\circle{40}}
   \put(-20,000){\line(-1,0){10}}
   \put(020,000){\line(+1,0){10}}
   \put(-10,-17){\line(0,+1){34}}
   \put(+10,-17){\line(0,+1){34}}
   \put(-11,-40){\text{\LARGE(c)}}
}}
\put(250,100){\makebox(0,0)[lb]{
   \put(000,000){\circle{40}}
   \put(-20,000){\line(-1,0){10}}
   \put(+20,000){\line(+1,0){10}}
   \put(000,000){\line(0,-1){20}}
   \put(000,000){\line(+5,+3){17}}
   \put(000,000){\line(-5,+3){17}}
   \put(-12,-40){\text{\LARGE(d)}}
}}
\put(320,100){\makebox(0,0)[lb]{
   \put(000,000){\circle{40}}
   \put(-20,000){\line(-1,0){10}}
   \put(+20,000){\line(+1,0){10}}
   \put(-14,014){\line(+1,-1){28}}
   \qbezier(-14,-14)(-08,-08)(-02,-02)
   \qbezier(+14,+14)(+08,+08)(+02,+02)
   \qbezier(-02,-02)(-04,004)(+02,002)
   \put(-10,-40){\text{\LARGE(e)}}
}}
\put(030,10){\makebox(0,0)[lb]{
   \put(000,000){\line(+1,0){40}}
   \put(000,000){\line(+3,+5){20}}
   \put(040,000){\line(-3,+5){20}}
   \put(000,000){\line(-5,-3){10}}
   \put(040,000){\line(+5,-3){10}}
   \put(020,033){\line(0,+1){10}}
   \put(010,017){\line(+1,0){20}}
   \put(009,-20){\text{\LARGE(f)}}
}}
\put(115,10){\makebox(0,0)[lb]{
   \put(000,000){\line(+3,+5){20}}
   \put(040,000){\line(-3,+5){20}}
   \put(000,000){\line(-1,-1){10}}
   \put(040,000){\line(+1,-1){10}}
   \put(020,033){\line(0,+1){10}}
   \put(000,000){\line(+5,+3){29}}
   \put(040,000){\line(-5,+3){17}}
   \qbezier(016,014)(013,016)(010.5,017.5)
   \qbezier(016,014)(024,018)(022.5,010.5)
   \put(009,-20){\text{\LARGE(g)}}
}}
\put(200,10){\makebox(0,0)[lb]{
   \put(000,000){\line(+1,0){40}}
   \put(000,000){\line(0,+1){40}}
   \put(040,000){\line(0,+1){40}}
   \put(000,040){\line(+1,0){40}}
   \put(000,000){\line(-1,-1){10}}
   \put(000,040){\line(-1,+1){10}}
   \put(040,000){\line(+1,-1){10}}
   \put(040,040){\line(+1,+1){10}}
   \put(020,040){\line(+1,-1){20}}
   \put(009,-20){\text{\LARGE(h)}}
}}
\put(285,10){\makebox(0,0)[lb]{
   \put(000,000){\line(+1,0){40}}
   \put(000,000){\line(0,+1){40}}
   \put(040,000){\line(0,+1){40}}
   \put(000,040){\line(+1,0){40}}
   \put(000,000){\line(-1,-1){10}}
   \put(000,040){\line(-1,+1){10}}
   \put(040,000){\line(+1,-1){10}}
   \put(040,040){\line(+1,+1){10}}
   \put(020,040){\line(0,-1){40}}
   \put(011,-20){\text{\LARGE(i)}}
}}
   \end{picture}
\end{center}
\phantom{ }\vspace{7truecm}\phantom{ } 
\caption{}
 \label{figureone}
 \end{figure}
}
\newcommand\mytoday{\number\day\space \ifcase\month\or
  January\or February\or March\or April\or May\or June\or
    July\or August\or September\or October\or November\or December\fi
      \space\number\year}

\def\eqref#1{Eq.(\ref{#1})}
\def\eqrefb#1#2{Eqs.(\ref{#1})-(\ref{#2})}
\def\itref#1{(\ref{#1})}

\begin{document}
\title{\vspace{1cm} \TITLE }
\author{S. Laporta\thanks{{E-mail: \tt laporta{\char"40}bo.infn.it}} \\ 
 \hfil \\ {\small \it Dipartimento di Fisica, Universit\`a di Bologna, }
 \hfil \\ {\small \it Via Irnerio 46, I-40126 Bologna, Italy} 
 }
\date{}
\maketitle
%\vspace{-7.5cm} \hspace{12.5cm} {\mytoday} \vspace{+7.5cm}
\vspace{-7.5cm} \hspace{12.5cm} {April 2000} \vspace{+7.5cm}
\vspace{1cm}
\begin{abstract}
In this paper we describe a new method of calculation of master integrals
based on the solution of systems of difference equations in one variable.
An explicit example is given, and the generalization to arbitrary diagrams
is described. As example of application of the method, 
we have calculated the values of master integrals for 
single-scale massive 
three-loop 
vacuum diagrams, three-loop self-energy diagrams, 
two-loop vertex diagrams and two-loop box diagrams.
\end{abstract}

\vspace{1.5cm}
PACS number(s):
\par  11.15.Bt General properties of perturbation theory
\par  02.90.+p Other topics in mathematical methods in physics
\par
  
\pagenumbering{roman}
\setcounter{page}{0}
\vfill\eject 
\pagenumbering{arabic}
\setcounter{page}{1}
In this paper we describe a new method of calculation of master integrals
based on solution of difference equations.
We introduce our method immediately by means of an example. 
Let us consider the integral 
\begin{equation}\label{inte0}
J(x)=\int \dfrac{\dk{}}{(k^2+1)^x}\ ,
\end{equation}
where $\dk{}=d^D k/\pi^{D/2}$. 
By using integration-by-parts\cite{Tkachov,Tkachov2},
one finds the recurrence relation 
\begin{equation}
\label{equ1den}
(x-1) J(x)-(x-1-D/2)J(x-1)=0\ .
\end{equation}
This recurrence relation is
a first-order linear \emph{difference equation}
with polynomial coefficients for the function $J$.
A solution of \eqref{equ1den} can be written
in the form of an expansion in factorial series
\begin{multline}\label{remfac}
J(x)=\mu^x \sum_{s=0}^\oo\frac{a_s \Gamma(x+1)}{\Gamma(x-\K+s+1)} \\
    = \mu^x\frac{\Gamma(x+1)}{\Gamma(x-\K+1)}\left(
         a_0+\frac{a_1}{x-\K+1} +\frac{a_2}{(x-\K+1)(x-\K+2)} +\ldots 
\right)\ ,
\end{multline}
where $\mu$, $\K$, and $a_s$ are 
to be 
determined.
By substituting the expansion \itref{remfac}
in the difference equation with the help of the Boole's operators\cite{Milne}
$\PI$ and $\RHO$, which allow one to rewrite \eqref{remfac} as a 
series of powers of $\RHO$, one finds the values $\mu=1$ and $\K=-D/2$, 
and the recurrence relation between the coefficients $a_s$ 
\begin{equation}\label{recu0}
s a_s= a_{s-1}(s+D/2)(s+D/2-1) \ .
\end{equation}
The first coefficient, $a_0$, 
can be determined by comparing the large-$x$ behaviour of the
series, $J(x)\approx \mu^x x^{\K} a_0 =x^{-D/2} a_0$,
with the large-$x$ behaviour of the integral \itref{inte0} 
\begin{equation}\label{inter1}
\int \dfrac{\dk{}}{(k^2+1)^x}\approx \int \dk{} e^{-x k^2}= x^{-D/2}\ ,
\end{equation}
from which one infers $a_0=1$.
For large $s$ the coefficients $a_s$ grow approximately as $s! s^{D-1}$,
similarly to coefficients of asymptotic expansions in powers of $1/x$,
but with the big difference that the factorial growth is compensated by
the gamma function in the denominator of \eqref{remfac},
so that the generic term of the sum behaves as $s^{D/2-1-x}$
and the series turns out to be \emph{convergent} for $x>D/2$.
This is a general feature of expansions in factorial series.
Furthermore,
by using the recurrence relation \itref{equ1den} we can evaluate $J(x)$
for values of $x$ outside 
the domain of convergence of the expansion \itref{remfac}.
Let us now consider the on-mass-shell integral
\begin{equation}\label{inte1}
I(x)=\int \dfrac {\dk{}}{(k^2+1)^x(k^2-2 p\cdot k)} \ , \quad p^2=-1 \ .
\end{equation}
Combining identities obtained by integration-by-parts one finds the 
recurrence relation
\begin{multline}\label{diffequ1}
( x - D )I(x - 2) + ( 2x - D - 1 )I(x - 1) -3(x-1)I(x) =
(1 -D/2)J(x - 1) \ ,
\end{multline}
where the function \itref{inte0} appears in the right-hand side.
This is a nonhomogeneous difference equation of second order. 
The general solution of this equation has the form
\begin{equation}\label{solgeni}
I(x)=\C_{1}I_{1}(x) + \C_{2}I_{2}(x) +I_3(x)\ ,
\end{equation}
where $I_1$ and $I_2$ are solutions of the homogeneous equation
(obtained by setting the right-hand side of \eqref{diffequ1} to zero),
$\C_1$ and $\C_2$ are arbitrary constants,
and $I_3$ is a particular solution of
the nonhomogeneous equation.
As before, we look for solutions in the form of expansions in factorial
series 
\begin{equation}\label{expfact}
I_j(x)=\mu_j^x 
    \sum_{s=0}^\oo\frac{b_s^{(j)} \Gamma(x+1)}{\Gamma(x-\K_j+s+1)}\ , \quad 
j=1,2,3\ ,
\end{equation}
where we set $b_0^{(1)}=b_0^{(2)}=1$.
By substituting the expansions \itref{expfact} and \itref{remfac}
in the difference equation one finds the values 
\begin{alignat*}{3}
\mu_1=&1,        \qquad&\mu_2=&-1/3,      \qquad& \mu_3=&1, \\
 \K_1=&-D/2+1/2, \qquad& \K_2=&-D/2+1/2 , \qquad&  \K_3=&-D/2\ ,
\end{alignat*}
and the recurrence relations between the coefficients $b_s^{(i)}$ 
\begin{equation}\label{recu1}
4s b_s^{(1)} -(\sD-1/2)(7\sD-D+1/2)b_{s-1}^{(1)}+3(\sD -3/2)(\sD-1/2)^2
b_{s-2}^{(1)}=0\ ,
\end{equation}
\begin{equation}\label{recu3}
(4s+2)b_s^{(3)} -\sD(7\sD-D+4)b_{s-1}^{(3)}+3\sD^2(\sD-1) b_{s-2}^{(3)}=
(D/2-1)(a_s-\sD a_{s-1}) \ ,
\end{equation}
 where $\sD = s+D/2-1$, and $a_i=b_i=0$ for $i<0$.
As we will see immediately, the recurrence relation for $b_s^{(2)}$ is not needed.
The constants $\C_1$ and $\C_2$
can be determined by comparing the large-$x$ behaviour of the series,
$I_1(x)\approx x^{-D/2+1/2}$,
$I_2(x)\approx (-1/3)^x x^{-D/2+1/2}$ and
$I_3(x)\approx (1/2-D/4) x^{-D/2}$,
with the large-$x$ behaviour of $I(x)$  
\begin{equation}\label{inter2}
I(x) \approx \int  \dfrac{\dk{} e^{-x k^2}}{k^2-2p \cdot k} 
\approx\dfrac{\sqrt{\pi}}{2}  x^{-D/2+1/2}   \ .
\end{equation}
One finds that $\C_1=\sqrt{\pi}/2$ and $\C_2=0$,
therefore the solution $I_2$ can be discarded.
A numerical value of $\C_1$ may be alternatively obtained from \eqref{solgeni}
evaluated at $x=0$, in the form $\C_1= (J(1)-I_3(0))/I_1(0)$.

Values of $J(x)$, $\C_1I_1(x)$, $I_3(x)$ and $I(x)$ calculated for $D=4-2\e$ are shown in
Table~\ref{tableij},
with the first two terms of the expansion in $\e$;
we also show the number of terms of the factorial series
needed to obtain the results with 16 exact digits.
Values for $x\le 2$ are obtained by using the recurrence relations
\itref{equ1den} and \itref{diffequ1}.
We see that for small $x$ the convergence is slow, so that
it is convenient to evaluate the series for $\bar x \gg 1$,
and to use repeatedly the recurrence relations
in order to obtain $J$ or $I$ for the given value of $x$;
note, however, that the recurrence relation \itref{diffequ1}
is unstable so that each application
increases the error on $I(x)$ of a factor $|\mu_1/\mu_2|=3$,
and therefore a value of $\bar x$ too large is not convenient.

We mention here that there is 
an alternative way to represent the functions $J$, $I_1$ and $I_3$
by using the Laplace's transformation\cite{Milne}:
 \begin{equation}\label{inint}
 J(x)=\int_0^1 dt \; t^{x-1} w(t) \ , \qquad I_j(x)=\int_0^1 dt \; t^{x-1}
 v_j(t)\ , \quad j=1,3 \ ,
 \end{equation}
 where $w$ and $v_j$ satisfy the differential equations
 \begin{equation}\label{equde1}
 t(1-t) w' + (D/2-t) w =0 \ ,
 \end{equation}
 \begin{equation}\label{equde3}
 %( 3t^3 - 2t^2 - t ) v_3'(t) + (3t^2- t(D-1)+2-D) v_3(t) = (1-D/2) t w(t)
 t(t-1)(3t+1)v_j' + (3t^2- t(D-1)+2-D) v_j = \delta_{j3}(1-D/2) t w\ ,
  \quad j=1,3\ .
 \end{equation}

Now we want to generalize the method used above to the calculation
of master integrals of arbitrary diagrams.
Let us consider a diagram with $\NK$ loops, $\ND$ internal lines
and  $L$ master integrals. Defined a particular order of the
denominators $D_1, \ldots, D_\ND$,
we group the master integrals
in different sets, such that each set contains
the master integrals $B_{ml}$ which have $D_m$ as first denominator:
\begin{equation}
B_{ml}=
\int {\dk1 \dots \dk{\NK}}
\dfrac
{N_{ml}}
{D_{m} D_{m+i_1} D_{m+i_2} \ldots D_{m+i_n}}\ , 
\quad
  0<i_1<i_2<\ldots<i_n \ ,
\end{equation}
where
$ m=1, \ldots, \ND-\NK+1$,
the indices $i_1, \ldots, i_n$ depend on $m$ and $l$,
and the numerators $N_{ml}$ in general contain  products of powers of some
irreducible scalar products involving the loop momenta,
which cannot be simplified by algebraic identities with one of denominators.
Integrals $B_{m1}, B_{m2}, \ldots$,
are ordered by increasing number of denominators
and, for integrals with the same $n$, 
by increasing sum of powers of scalar products contained in $N_{ml}$.

Changing the exponent of $D_m$ from 1 to $x$ 
in each master integral, we define the ``master functions''
\begin{equation}\label{definitionu}
I_{ml}(x)=
\int {\dk1 \dots \dk{\NK}} \  
\dfrac
{N_{ml}}
{D_{m}^x D_{m+i_1} D_{m+i_2} \ldots D_{m+i_n}}\ . 
\end{equation}
In general, for each possible choice of $D_m$ equal to one of the
denominators,
a different function $I_{ml}$ is defined; 
but for $x=1$ they all must have the same value $I_{ml}(1)=B_{ml}$.
This fact is particularly useful for checking the consistency
of the calculations. 

By combining by suitable algorithms\cite{Lapold}
the identities obtained by integration-by-parts,
for each value of $m$ we build a
system of linear difference equations in triangular form 
\begin{equation}\label{diffu1}
\sum_{i=0}^{\R_l} p_{il}(x) I_{ml}(x+i) = 
\sum_{k=1}^{l-1} \sum_{j=0}^{Q_{lk}} q_{jkl}(x) I_{mk}(x+j) \ , \quad
 l=1,\ldots,L_m \ ,\quad 
\end{equation}
satisfied by the functions $I_{mj}$,
with polynomial coefficients $p_{il}$ and $q_{jkl}$,
where the right-hand side of the $l$-th equation contains the functions
from $I_{m1}$ to $I_{m,l-1}$.
The triangular structure is particularly useful for simplifying 
the numerical solution:
the equations are solved one at a time, for $l=1, 2, \ldots, L_m$;
when the equation for $I_{ml}$ has to be solved,
all the functions $I_{m1}, \ldots, I_{m,l-1}$ in the right-hand side
are already known.
Therefore each single equation is similar to \eqref{equ1den} or
\eqref{diffequ1}, 
with the difference that, for 
integrals more complicated than \eqref{inte1},
the order of the equation and the degree of
polynomials of coefficients are generally larger, 
and the right-hand side contains the sum of several contributions.
The solution of each equation of the system \itref{diffu1} has the form
\begin{equation}\label{solg}
I_{ml}(x)=\sum_{j=1}^{\R_l} \C_j \IOMOG_j(x) + \INOMOG(x)  
\end{equation}
for integer values of $x$.
Each solution  $\IOMOG_j(x)$ of the homogeneous equation,
and the particular solution $\INOMOG(x)$ of the nonhomogeneous equation  are expanded in
factorial series
\begin{equation}\label{remfac2}
\IOMOG_j(x)=\mu_j^x \sum_{s=0}^\oo\frac{a_s^{(j)} 
\Gamma(x+1)}{\Gamma(x-\K_j+s+1)}\ ,
\quad
\INOMOG(x)=\sum_{j=1}^S (\mu^{NH}_j)^x \sum_{s=0}^\oo\frac{a_s^{(j,NH)} 
\Gamma(x+1)}{\Gamma(x-\K_j^{NH}+s+1)}\ .
\end{equation}
Values of $\mu_j$, $\K_j$ and recurrence relations between coefficients
$a_s^{(j)}$ and $a_s^{(j,NH)}$ 
can be found by using the operator method mentioned in the previous
example; the values of $\mu^{NH}_j$ and $\K_j^{NH}$ 
come from the already known expansions in factorial series 
of the nonhomogeneous term of \eqref{diffu1}.
The recurrence relations between 
$a_s^{(j)}$ and $a_s^{(j,NH)}$ 
are similar to
\eqrefb{recu1}{recu3},
but in general they are of higher order,
and the coefficients of the expansions of the
functions $I_{m1}, \ldots, I_{m,l-1}$ appear in 
the right-hand side of the recurrence relations for $a_s^{(j,NH)}$.

A comparison between
the large-$x$ behaviours of the factorial series \itref{remfac2}
and 
of the integral $I_{ml}(x)$
allows one
to determine which constants $\C_j$ are zero
(because the corresponding values of $\mu_j$ and $K_j$ are not compatible) 
and which must be calculated.

Let us now consider the large-$x$ behaviour of the
master function $I_{ml}(x)$.
Choosing the momentum routing such that $D_m=k_1^2+m_1^2$, 
introducing hyperspherical polar coordinates for the integration over $k_1$
and separating the angular and radial part,
\eqref{definitionu} becomes
\begin{equation}\label{intk2}\label{int00}
I_{ml}(x) = \dfrac{1}{\Gamma(D/2)}
  \int_{\ell}  \dfrac{d k_1^2 \;(k_1^2)^{D/2-1}} {(k_1^2+m_1^2)^x}
     f(k_1^2) \ ,
\end{equation}
where $f$ is given by
\begin{equation}\label{definf}
 f(k_1^2)=
     \dfrac{1}{\Omega_D}
 \int d\Omega_{D}(\hat k_1) \; 
 \int  \  
\dfrac{\dk2 \; \dots \dk{\NK}\ N_{ml}}{\displaystyle D_{m+i_1} \ldots  D_{m+i_n}} \  \ .
\end{equation}
The path of radial integration $\ell$ is the straight line $[0,\oo]$
for euclidean integrals.
For non-euclidean integrals,
in order to ensure a correct analytical continuation,
the path $\ell$ must be deformed\cite{Levinehyp1,Levinehyp2},
turning around some singularities
when the external momenta get over some values of threshold
in the non-euclidean region;
for example, considering a one-loop self-energy integral with $m_1=m_2=1$, 
the path must be deformed
if $p^2<-1$. The integral \itref{intk2} for large $x$ receives contributions
from a neighbourhood of the origin 
and from each singularity of the function $f(k_1^2)$
which is on the path $\ell$, turning points included.

We consider here only the case where
the path $\ell$ is not deformed and all the masses are non-zero,
so that there is only the contribution from the origin.
More general cases will be discussed elsewhere.
Under this assumption and if, for simplicity, 
$f(0)$ has a non-zero finite value,
the large-$x$ behaviour of the master function is given by 
\begin{equation}\label{expkf00}
I_{ml}(x)
\approx 
(m_1^2)^{D/2-x} x^{-D/2} f(0) \ .
\end{equation}
Only the solutions $\IOMOG_j(x)$ with $\mu_j=1/m_1^2$ and
$K_j+D/2$ equal to a non-negative integer have a large-$x$ behaviour
compatible with \eqref{expkf00}
(often there is only one such solution).

$f(0)$ is expressed by an integral
over $k_2,\ldots,  k_\NK$, belonging to a diagram with \emph{one loop less},
obtained by setting $k_1=0$ in the original integral; it may be calculated 
by inserting an exponent in a denominator and building and solving
new difference equations.
If the integral \itref{definitionu} contains $N$ on-mass-shell denominators
$D_{m+i_1}=k_1^2-2 p_1 \cdot k_1$, 
\ldots, 
$D_{m+i_N}=k_1^2-2 p_N \cdot k_1$,
the large-$x$ behaviour of $I_{ml}$ becomes  
\begin{equation}\label{expkf01}
I_{ml}(x) \approx (m_1^2)^{D/2-N/2-x}  x^{-D/2+N/2}
\dfrac{\tilde f(0)}{2}
\int\dfrac{[d^N q]}{(q^2-2 p_1\cdot q)\cdots (q^2-2 p_N\cdot q)}
\ ,
\end{equation}
\begin{equation}
\tilde f(k_1)=\int 
\dfrac
{\dk2 \cdots \dk{\NK} \ N_{ml}}
{D_{m+i_{N+1}} \cdots D_{m+i_n}}\ ,
\end{equation}
if $\tilde f(0)$ has a non-zero finite value.
Note that the number of dimensions of the integral in \eqref{expkf01}
is the number of on-mass-shell denominators.
An additional relation between constants $\C_j$  may be found by evaluating
\eqref{solg} at $x=0$, as shown in the example.

Once the values of the arbitrary constants $\C_j$ are obtained,
the factorial series \itref{remfac2} must be summed.
If the coefficients $a^{(j)}_s$ or $a^{(j,NH)}_s$ grow as $s! B^s$ for large
$s$,
with $|B|>1$, the corresponding factorial series does not converge for any $x$.
This may occur for particular values of masses and external momenta of the 
master integral. In this case we resort to the Laplace's transformation 
mentioned in \eqref{inint}, by re-expressing \eqref{remfac2} as 
 \begin{equation}\label{inint2}
 \IOMOG_j(x)=\int_0^{\mu_j} dt \; t^{x-1} v^{HO}(t) \ , \qquad \INOMOG(x)=
 \sum _{j=1}^S \int_0^{\mu^{NH}_j}  dt \; t^{x-1}
 v^{NH}_j(t)\ , 
 \end{equation}
where $v^{HO}$ and $v^{NH}_j$ satisfy respectively homogeneous and
nonhomogeneous differential equations obtained by substituting these integral
representations in the difference equation; the differential equations are
conveniently solved by expanding in power series the solutions around a number
of points selected on the interval of integration.

Now we consider the application of our method to the calculation of master
integrals.
Considered the number of master integrals and the complexity
of the systems of difference equations for diagrams with two or more loops,
the use of an automated tool is necessary.
We have written on purpose a comprehensive program, called $\SYS$,
which automatically
determines the master integrals of a diagram, builds the systems
of difference equations and 
solves 
the systems 
numerically
by using expansions in
factorial series or integral representations.
The program contains a simplified symbolic manipulator, used for solving
the systems of integration-by-parts identities.
Expansion in $\e$ of the results is obtained
by expanding everywhere in $\e$ and
by manipulating
truncated series
by means of the arithmetical routines implemented in the program;
in this way all coefficients of expansions in $\e$
are obtained in numerical form, divergent terms
included.
Very high-precision results (even 50-100 digits) can be easily obtained, 
as the calculation of master integrals 
is reduced to the sum of convergent factorial or power series in one variable.

For illustrating the power of our method,
we have calculated the master integrals of the diagrams shown in
Fig.\ref{figureone} for $D=4-2\e$ 
in the single-scale case, with all masses equal to one,
with all the external 
lines
on-mass-shell and,
for box diagrams, the Mandelstam variables set to $s=t=1$ and $u=2$.
The number of master integrals is 3, 3, 14, 21, 17, 6, 5, 11 and 11, respectively
for the diagrams (a)-(i)
(not counting subdiagrams which factorize, or with external lines united).
The systems of difference equations between master functions 
generated by the program are formed with
 44, 28, 245, 304, 362, 68, 81, 139 and  158 equations, respectively;
at present
the program makes no use of the symmetries due to the particular values of
masses and momenta in order to simplify or reduce the number of the equations.
Note that for each subdiagram there may be several different master functions,
 according to which denominator is raised to $x$.
Use of Laplace's transformation is required for obtaining the master integrals
(e), (g) and (i).
As an example, 
the calculation from scratch
of all the master integrals of the diagram (c),
by using a precision of 38 digits, 
requested about 128 hours of CPU time on a 133 MHz Pentium PC;
we stress that at this preliminary stage of development
we directed our efforts to devise tests and cross checks rather
than to speed up the program.

For brevity, we show here only
the values of the master integrals with numerator equal to one,
corresponding to the diagrams shown in figure,
leaving the list of the remaining master integrals
to a further more exhaustive paper\cite{Lapold}.
As usual, the results have been normalized
with the division by $\Gammae\equiv\Gamma(1+\e)$ raised to
the number of loops of the diagram.
\begin{multline}\label{2e}
I(\text a)\Gammae^{-3}=
 2.404113806319 \e^{-1} -10.03527847977 +35.94478903214 \e \\
 -119.1503507802 \e^2 +379.7433345095 \e^3 -1183.320931551 \e^4
   +O(\e^5)\;,
\end{multline}
\begin{multline}\label{res3d}
I(\text b)\Gammae^{-2}=
0.9236318265199 -1.284921671848 \e +2.689507626490 \e^2 \\
-5.338399227511 \e^3 +10.67136736912 \e^4 
   +O(\e^5)\;,
\end{multline}
\begin{multline}\label{res3t}
I(\text c)\Gammae^{-3}=
 0.2796089232826 -0.1380294113932   \e  +0.3194688268113 \e^2 \\
 -0.4399664109267 \e^3 +0.6650515012166 \e^4 
   +O(\e^5)\;,
\end{multline}
\begin{multline}
I(\text d)\Gammae^{-3}=
 0.1826272375392 -0.06746690965803  \e  +0.1865462420623 \e^2 \\
 -0.2498713405447 \e^3 +0.3796187113121 \e^4 
   +O(\e^5)\;,
\end{multline}
\begin{multline}
I(\text e)\Gammae^{-3}=
 0.1480133039584 -0.009263002043238 \e  +0.1053308537397 \e^2 \\
 -0.1224292041846 \e^3 +0.1898480457555 \e^4 
   +O(\e^5)\;,
\end{multline}
\begin{multline}
I(\text f)\Gammae^{-2}=
0.2711563494022 +0.1833941077514 \e +0.05375101058769 \e^2 \\
+0.01446103368419 \e^3 +0.000746187372276 \e^4 
   +O(\e^5)\;,
\end{multline}
\begin{multline}
I(\text g)\Gammae^{-2}=
 0.173896742268 +0.1816664876962 \e +0.04440899181832 \e^2 \\
 +0.02231547385785 \e^3 -0.003079810479797 \e^4 
   +O(\e^5)\;,
\end{multline}
\begin{multline}
I(\text h)\Gammae^{-2}=
0.1723367907503 +0.2679578491711 \e +0.13552112755141 \e^2 \\
+0.04468531532833 \e^3 +0.008430602827459 \e^4 
   +O(\e^5)\;,
\end{multline}
\begin{multline}\label{res4h}
I(\text i)\Gammae^{-2}=
0.1036407209893 +0.2142416932987 \e +0.14046068671363 \e^2 \\
+0.04437197236388 \e^3 
   +O(\e^4)\;.
\end{multline}
Of these results, only first two terms of \eqref{2e} and the first term
of \eqref{res3d} were already known\cite{2e,3abcd2}. Remaining terms and other
results are new.

In conclusion,
in this work we have established a connection between recurrence relations
obtained by integration-by-parts and  difference equations.
Using theory of difference equation
we have developed a new method of calculation of master integrals
which can be applied to diagrams with any topology and any number of loops.
High-precision results expanded at will in $\e$ can be obtained easily.
We have demonstrated the actual validity of the method,
as first application, 
by calculating master integrals in the single-scale case.
Application of the method to different cases,
for example diagrams with different masses, external momenta over threshold,
or zero masses, 
which involves 
the solution of 
higher-order 
difference equations 
or more 
laborious 
determinations of large-$x$ behaviours,
will be described in future papers.

\section*{Acknowledgement}
The author wants to thank E. Remiddi and M. Caffo for
useful discussions and encouragement in the very early stage of this work.

\vfill\eject 
\pagenumbering{roman}
\setcounter{page}{1}
\phantom{.}\vspace{5cm}
\section*{Figure Captions}
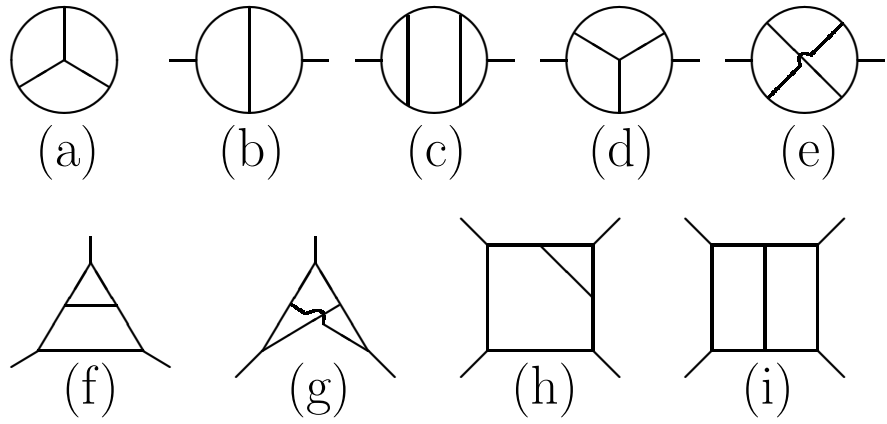
\par\noindent Figure 1: \CAPTIONFIG
\phantom{.}\vspace{12cm}\IDENTIFY
\vfill\eject 
\phantom{.}\vspace{5cm}
\TableOne
\phantom{.}\vspace{7cm}\IDENTIFY
\vfill\eject 
\phantom{.}\vspace{1cm}
\FigureOne
\phantom{.}\vspace{1cm}\IDENTIFY
\end{document}